# Ballistic transport in nanodevices based on single-crystalline Cu thin film

Yongjin Cho[1], Su Jae Kim[2], Min-Hyoung Jung[3], Yousil Lee[4], Hu Young Jeong[5], Young-Min Kim[3], Hu-Jong Lee[1], Seong-Gon Kim[6]*, Se-Young Jeong[7,8,9]* & Gil-Ho Lee[1,10]*

[1]Department of Physics, Pohang University of Science and Technology, Pohang, Republic of Korea.

[2]Crystal Bank Research Institute, Pusan National University, Busan 46241, Republic of Korea.

[3]Department of Energy Science, Sungkyunkwan University, Suwon 16419, Republic of Korea.

[4]Copper Innovative Technology (CIT) Co. Busan 46285, Republic of Korea

[5] Graduate School of Semiconductor Materials and Devices Engineering, Ulsan National Institute of Science and Technology, Ulsan 44919, Republic of Korea

[6]Department of Physics and Astronomy, Mississippi State University, Mississippi State, MS 39762, USA.

[7]Department of Physics, Korea Advanced Institute of Science and Technology (KAIST), Daejeon 34141, Korea

[8]Gordon Center for Medical Imaging, Department of Radiology, Massachusetts General Hospital and Harvard Medical School, Boston, MA, 02114, USA.

[9]Department of Optics and Mechatronics Engineering, Engineering Research Center for Color-Modulated Extra-Sensory Perception Technology, Pusan National University, Busan 46241, Republic of Korea.

[10]Asia Pacific Center for Theoretical Physics, Pohang, Republic of Korea.

*Corresponding authors. E-mail:
sk162@msstate.edu (S.-G.K.);
sjeong9@kaist.ac.kr (S.-Y.J);
lghman@postech.ac.kr (G.-H.L)

**In ballistic transport, the movement of charged carriers is essentially unimpeded by scattering events. In this limit, microscopic parameters such as crystal momentum, spin and quantum phases are well conserved, allowing electrons to maintain their quantum coherence over longer distances. Nanoscale materials, like carbon nanotubes, graphene,**

**and nanowires, exhibit ballistic transport. However, their scalability in devices is significantly limited. While deposited metal films offer excellent scalability for nanodevices, achieving ballistic transport in these films poses a challenge due to their short electronic mean free path. Here, we investigated the electronic transport in cross-geometry devices fabricated with 90 nm-thick copper films without grain boundaries. We observed ballistic transport in devices with channel width smaller than 150 nm below 85 K by measuring negative bend resistance. Our findings would open the opportunity for probing intrinsic quantum properties of Cu, and for realizing scalable low-loss signal transmission and high-quality interconnects in semiconductor devices.**

**Introduction**

When the electronic mean free path is longer than the device size, electrons exhibit ballistic transport without being significantly affected by scatterings with phonons, rough surfaces, impurities, and crystalline defects such as grain boundaries (GBs), which are the boundaries between two adjacent grains with different crystal orientations. Ballistic transport can reveal the intrinsic quantum properties of solid-state materials, as the quantum information of electrons—such as crystal momentum, spin, and quantum phase—is well conserved[1,2,3,4]. In recent decades, there has been enormous research about quantum properties of normal metals, such as electronic band structure, Fermi surface topology, electronic conductivity, magnetoresistance[5,6,7,8,9], etc. Among various metals, copper (Cu) has a long history of being used in electronic circuits due to its unparalleled combination of conductivity, reliability, and versatility. It remains an important material in the electronics industry, used for everything from high-speed data-processing cables to interconnects in advanced semiconductor devices. Electronic band structure, Fermi surface topology and Fermi velocity of bulk Cu are well understood by de Haas-van Alphen effect[10], galvanomagnetic phenomena[11,12], photoemission[13], cyclotron resonance[14], anomalous skin effect[15], and magnetoacoustic[16] measurements. However, growth of atomically uniform and ultraflat Cu thin films without crystalline defects such as GBs has been technically challenging, and the ballistic transport in nanostructures based on Cu thin films has not yet been achieved, limiting the ability to exploit their intrinsic quantum properties.

Here, we report experimental observation of ballistic transport in the nanoscale devices fabricated with atomically flat single-crystalline Cu(111) thin film (SCCF) (Supplementary Fig. 1), which is grown by atomic sputtering epitaxy (ASE) technique[17], which offers a solution to the technical challenges in conventional metal-film deposition, such as surface oxidation, degradation of electrical performance, electromigration resulting from defects such as GBs and impurities[18]. We observed negative bend resistance in cross-geometry devices, indicative of ballistic transport. The electronic mean free path of SCCF with a thickness of approximately 90 nm was estimated to be 150 nm below 85 K (Supplementary Fig. 2). Our SCCF can facilitate studies on quantum nature of Cu thin films, including topological properties[19,20,21], quantum Hall effect, hydrodynamic electron transport[22], and phase-coherent quantum interference, which can be extended to studies on other metallic thin films. In addition, it suggests the potential for applications in quantum circuits, spintronic devices, and Cu interconnect technology[23,24,25].

# Results

## Ballistic transport in a SCCF

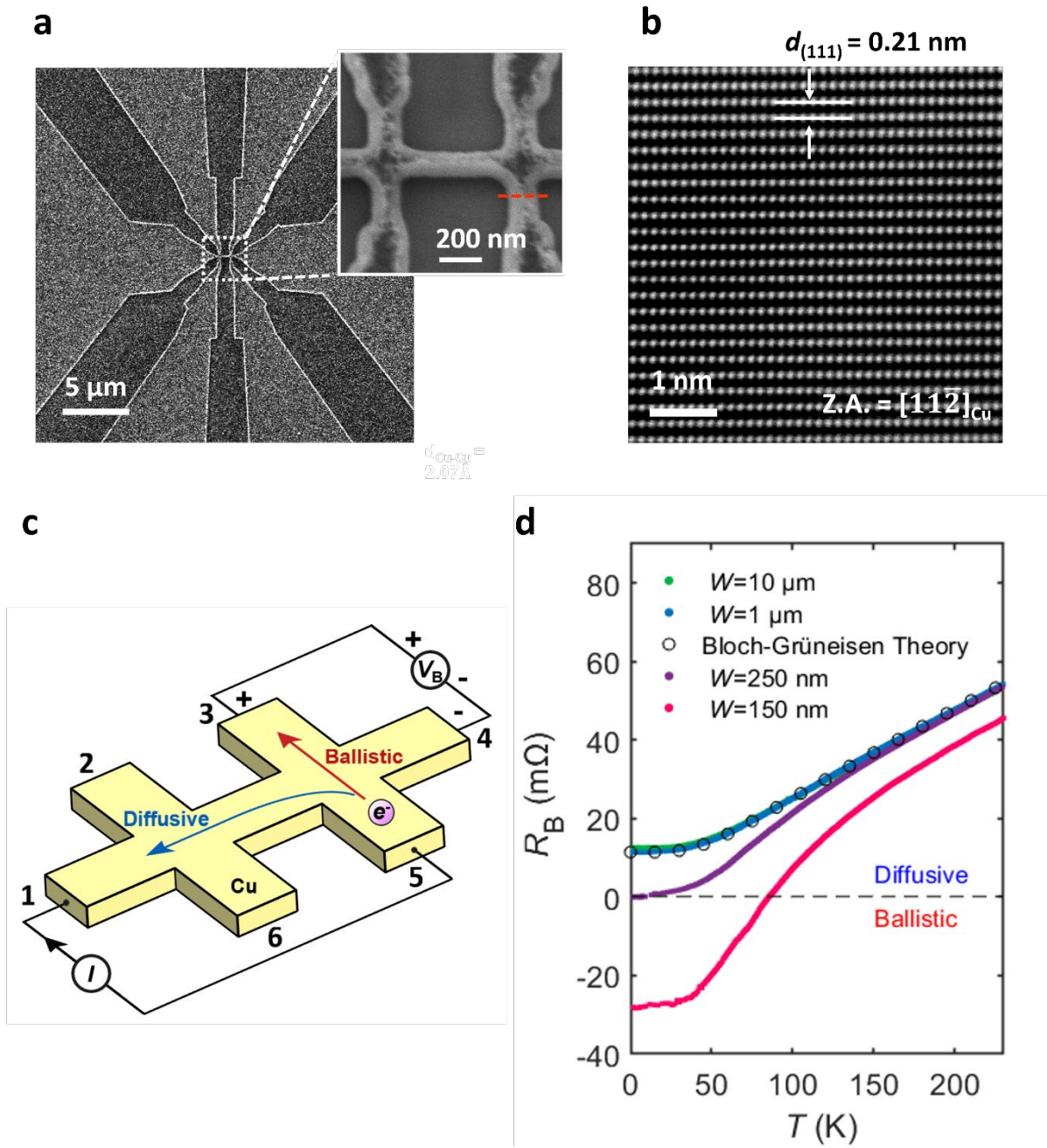

**Fig. 1 | Ballistic transport in a single-crystalline Cu(111) thin film (SCCF). a,** Scanning electron microscopy (SEM) image of a Hall bar device with a channel width (*W*) of 150 nm. The inset illustrates the enlarged SEM image of the marked area in **a**. **b,** Transmission electron microscopy (TEM) image showing the cross-sectional view of the Hall bar device, prepared

using focused ion beam (FIB) milling along the red line indicated in the inset of **a**. **c,** Schematic representation of the bend-resistance ($R_B$) measurement. Electron trajectories for diffusive and ballistic transport are indicated by blue and red arrows, respectively. **d,** Temperature ($T$) dependence of $R_B$ in Hall bars with $W$=10 μm, 1 μm, 250 nm, and 150 nm. Circle symbols depict the fitting of $R_B$ with $W$=1 μm and 10 μm using Bloch-Grüneisen function.

With standard electron beam lithography and Argon ion etching process, Hall bar-shaped devices with various channel widths ($W$) of 10 μm, 1 μm, 250 nm, and 150 nm were fabricated with a thickness ($t$) ~90 nm (Figs. 1a,b and Supplementary Fig. 3). A cross-sectional TEM image of the device reveals perfectly aligned Cu atoms with 2.07 Å spacing in the [111] direction of the thin film (Fig. 1b). This suggests that the high crystalline quality of the thin film remains intact even after the fabrication processes (Supplementary Fig. 4). The bend resistance ($R_B$= $V_B/I$) was measured in the cross configuration (Fig. 1c), where current ($I$) is injected from terminal 1 to terminal 5 and the bend voltage ($V_B$) is measured between terminals 3 and 4. In a diffusive transport regime where the electronic mean free path ($l_{mfp}$) is shorter than the device length scale, $R_B$ can be described by van der Pauw formula $R_B=(\rho \ln 2)/(t\pi)$ for symmetric cross geometry. In this regime, $R_B$ is expected to be positive[26] as the resistivity, $\rho$, of material is inherently positive. The temperature dependence of $\rho$ in diffusive metals is predominantly influenced by electron-acoustic phonon scattering, which follows the Bloch-Gruneisen formula (Eq. 1) derived from Boltzmann transport theory[27].

$$\rho(T) = \rho_0 + \alpha_{el-ph}\left(\frac{T}{\Theta_R}\right) \int_0^{T/\Theta_R} \frac{x^5}{(e^x-1)(1-e^{-x})} dx \quad \text{(Eq. 1)}$$

Here, $\rho_0$ represents the residual resistivity due to temperature-independent scatterings by defects such as GBs and impurities. $\alpha_{el-ph}$ represents the electron-phonon coupling strength and $\Theta_R$ is Debye temperature. As shown in Fig. 1d, the temperature dependence of $R_B$ for devices of $W$=10 μm and 1 μm well fits Eq. 1 with fitting parameters $\alpha_{el-ph}$=8.8×10⁻⁸ and $\Theta_R$=270 K, which implies a diffusive transport over the entire temperature range. Indeed, $\rho$ determined by van der Pauw formula $\rho=(t\pi R_B)/(\ln 2)$ in fully diffusive regime for $W$=10 μm gives $l_{mfp}$ of ~170 nm at 1.7 K by using Eq. 2, which is much shorter than the device size.

$$l_{mfp} = \frac{m^* v_F}{ne^2 \rho} \quad \text{(Eq. 2)}$$

However, $R_B(T)$ of devices of $W \leq$ 250 nm, which is comparable to or shorter than $l_{mfp}$, clearly deviates from the value expected from Bloch-Gruneisen formula. For the device of $W$=150 nm,

$R_B$ even shows negative values below 90 K. Such counter-intuitive behaviour of negative bend resistance has been observed in high-mobility two-dimensional electronic systems like GaAs/AlGaAs heterostructures and graphene[3,4,28] as a consequence of ballistic transport[29]. However, it has never been observed in deposited metallic films. In a ballistic regime, the electrons injected from terminal 5 can directly propagate to terminal 3 without any scattering as indicated by red arrow line in Fig. 1c. As electrons accumulate near terminal 3, a negative electrical potential builds up at terminal 3. This results in the decrease of $R_B$, and even leading to negative value of $R_B$. From the findings above, we confirm that the ballistic transport in SCCF, previously hidden due to electron scattering at GBs, has been uncovered, when GBs are completely eliminated.

**Structural defects in SCCF and polycrystalline Cu thin film (PCCF)**

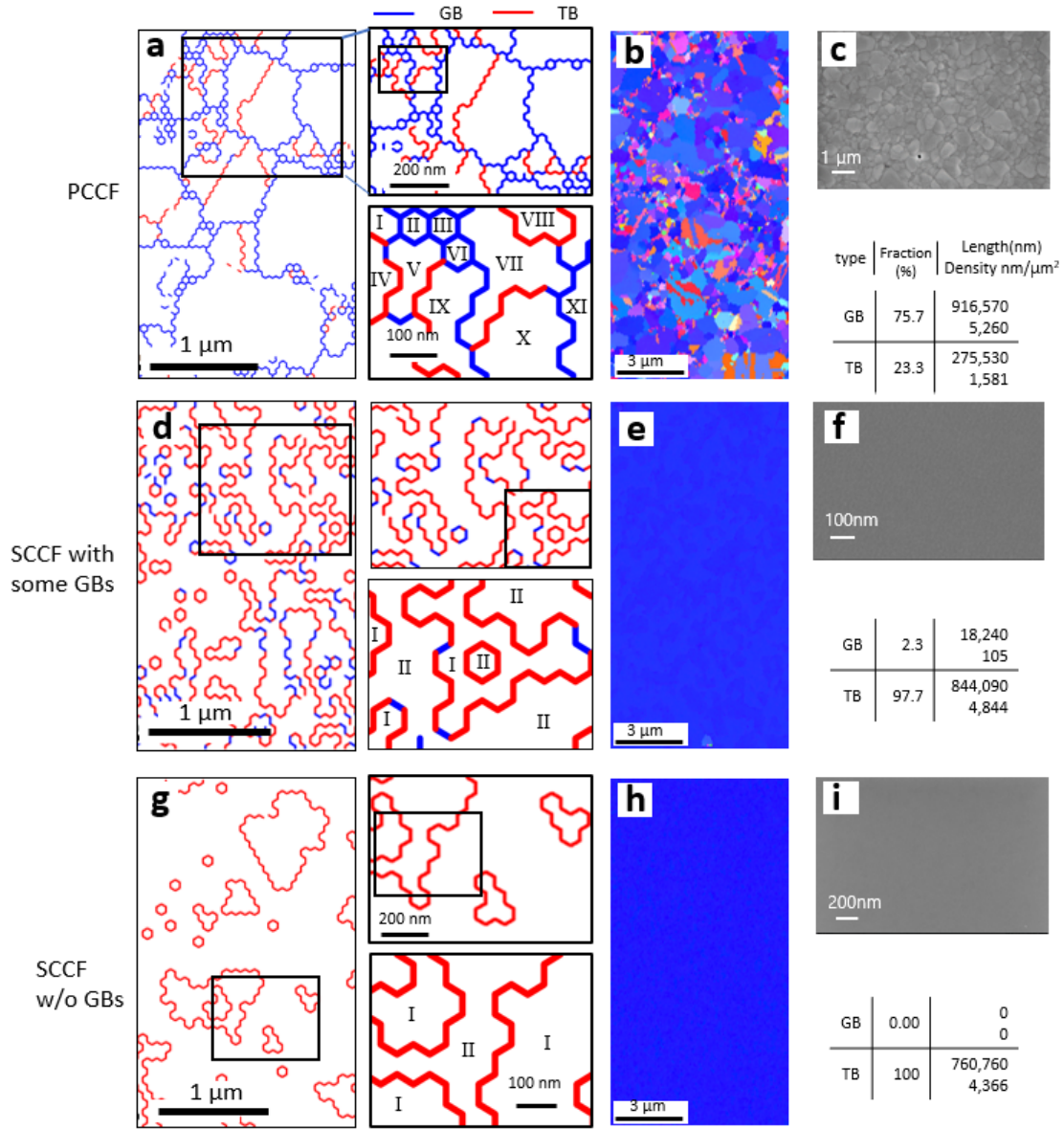

**Fig. 2 | Grain boundaries (GBs) and twin boundaries (TB) in polycrystalline Cu thin film (PCCF) and single-crystalline Cu(111) thin film (SCCF). a,** Misorientation line map of PCCF and enlarged images of marked area in each map. GBs and TBs are depicted by blue and red lines, respectively. For PCCF, misorientation-line distributions indicate the presence of both TBs and GBs. The Roman numerals in the bottom right panel represent the orientations of each grain. In the 2-inch wafer-sized PCCF, there are approximately trillions ($10^{12}$) different orientations[32]. **b,** Electron backscatter diffraction (EBSD) map of PCCF showing random alignment. **c,** SEM image of PCCF showing rough surface and GBs. **d,** Misorientation line map of a SCCF with a greatly reduced number of GBs and enlarged images of marked area in each map. **e,** EBSD map of **d** sample showing perfect alignment along the (111) plane. **f,** SEM image

of **d** showing high quality surface and no trace of GBs. **g,** Misorientation lines map of a SCCF without any GBs and enlarged images of marked area in each map. The Roman numerals in the bottom right panel represent the orientations of each grain. In the 2-inch wafer-sized SCCF, there are only two different orientations, which are stacked along ABC... and ACB... respectively. **h,** EBSD map of **g** sample showing perfect alignment along the (111) plane. **i,** SEM image of **g** showing high quality surface and no trace of GBs. Right upper panels of **a**, **d** and **g** are enlarged images of boxed area of left panels and right lower panels are enlarged images of boxed area of upper panels, respectively. Tables on the right of **b**, **e** and **h** give the fraction of GB and TB, and their density in the length (nm/□m$^2$) of each sample.

We analysed the crystallographic microstructure of SCCF as high crystalline quality of SCCF is essential for ballistic transport. In the thin film growth, the formation of GBs and TBs occurs due to lattice mismatch with the substrate. While the formation of TBs cannot be avoided, the formation of GBs can be minimized by utilizing a growth technique that considers extended atomic distance mismatch (EADM)[18]. The distribution of GBs and TBs was analysed for a PCCF (Fig. 2a, b and c), an SCCF with minimal GBs (Fig. 2d, e and f), and an SCCF without GBs (Fig. 2g, h and i) using misorientation line mapping, electron backscatter diffraction (EBSD) and SEM[30,31]. GBs and TBs are denoted by blue and red lines, respectively, in Fig. 2. The PCCF usually contains both GBs and TBs. The enlarged misorientation maps of boxed areas in a, d, and g left panels are shown in the right upper panels and those of boxed areas in upper panels are shown in the right lower panels. As shown in Fig. 2a, PCCF has lot of grains (Conventional PCCF is expected to have trillions of grains on the two-inch wafer[32] and each grain has a different orientation, as illustrated by the numbered indicators in the right lower panel of Fig. 2a. The Cu thin films grown by ASE show perfect alignment along the (111) plane (Fig. 2e) even though it has some GBs (Fig. 2d). GBs observed in the misorientation line map of Fig. 2d are very close to TBs and are not detected in EBSD map as different grains, because they are only slightly deviated the ideal TB condition of 60° (1-2°)[33]. In spite of that, those GBs can give an influence to the elecrtronic transport, thus we used only the SCCF without GBs for this study. A well-grown SCCF (Fig. 2g) is free of GBs, but consists of TBs only and exhibits only two orientations of I and II (Fig. 2i right lower panel). The crystallographic orientation of SCCF must follow one of ABC… and ACB... stacking and these two ORs satisfy symmetry operation by 60° rotation (Supplementary Fig. 5). The density of

TBs fluctuates across different areas, yet the overall density remains within same order of magnitude. In contrast to GBs, TBs have minimal effect on electrical resistivity for two primary reasons. Firstly, no charged defects are formed around TB, resulting in negligible potential variation near TB[34,35]. Secondly, well-matched Fermi surfaces of adjacent grains on either side of a TB reduce scattering of incident electron wave at the TB, because the Fermi surface of Cu remains unchanged with a 60° rotation about the (111) axis[27,33,35,36,37].

**Grain boundary (GB) dependence of electrical transport**

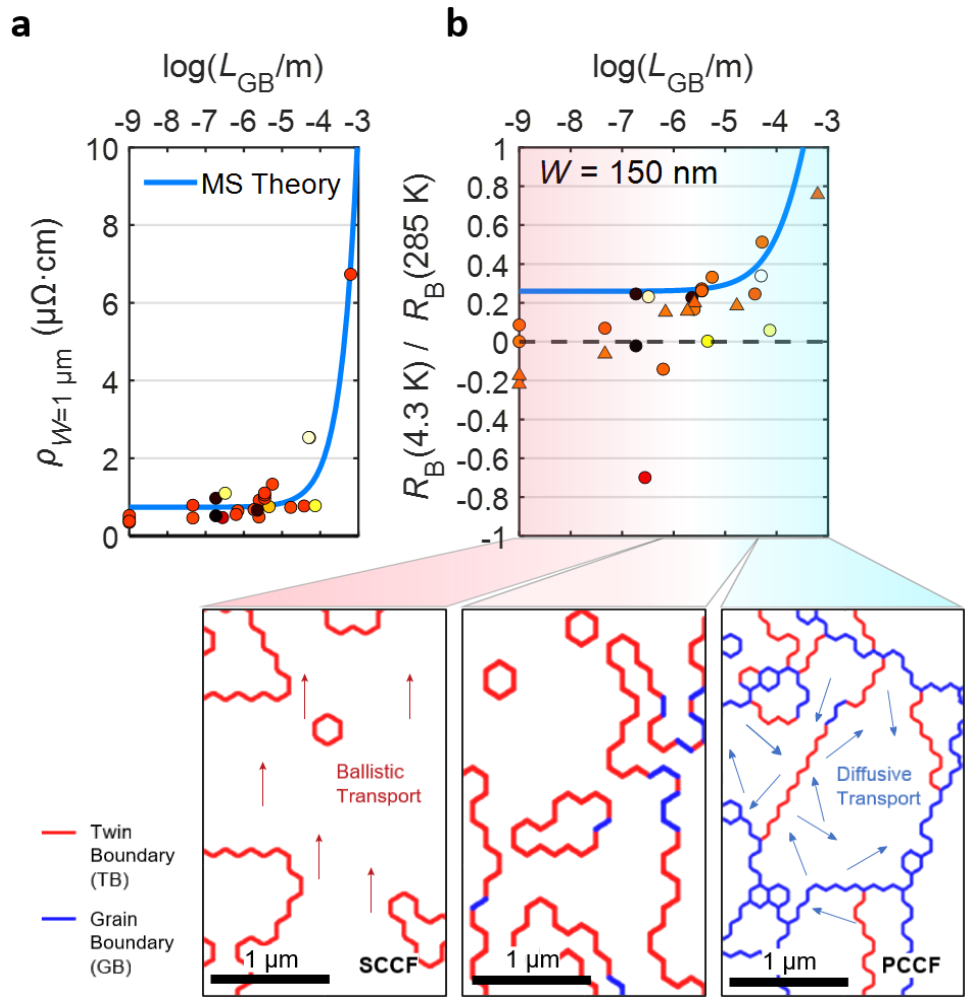

**Fig. 3 | GB dependence of electrical transport. a,** Longitudinal resistivity measured at 4.3 K as a function of GB length $L_{GB}$ for the device of width $W=1$ μm. Blue solid line represents the Mayadas and Shatzkes (MS) theory fit. **b,** Normalized bend resistance $R_B$ as a function of $L_{GB}$ for $W=150$ nm. Blue solid line represents the MS theory fit from **a**, circles and triangles represent the device without and with annealing processes, respectively. Colour of the symbols represent the thickness, $t_{Cu}$. Bottom panels represent misorientation line maps of single-

crystalline Cu(111) thin film (SCCF) without GBs (left), SCCF with some GBs (middle) and poly-crystalline Cu thin film (PCCF) (right).

Figure 3 shows the electronic transport behaviour with different GB length $L_{GB}$, which is the total length of line segment separating grains in misorientation line maps of the area of 7.7 μm by 22.7 μm measured by EBSD technique (Supplementary Fig.5)[30,31]. More GBs generally increase the resistivity by acting as scattering sites for electrons. Hence, as the grain size ($d$) decreases, i.e., as $L_{GB}$ increases, the resistivity contribution of GB increases. Mayadas and Shatzkes (MS) have studied more quantitative model to describe electron scattering at GBs[38,39,40]. According to MS model in Eq. 3, the resistivity decreases as the grain size ($d$) increases and increases with the probability of electron scattering at GB ($S$).

$$\frac{\rho_g}{\rho_i} = \left\{3\left[\frac{1}{3} - \frac{1}{2}\alpha + \alpha^2 - \alpha^3 \ln\left(1 + \frac{1}{\alpha}\right)\right]\right\} \quad \text{(Eq. 3a)}$$

$$\alpha = \frac{l_{MFP}}{d} \frac{S}{1-S} \quad \text{(Eq. 3b]}$$

Here, $\rho_g$ represents the resistivity resulting from the electron scattering at GBs and while $\rho_i$ represents the resistivity due to other causes including scattering with acoustic phonons, and defects such as impurities, and grain boundaries. As shown in Fig. 3a, $L_{GB}$ dependence of the resistivity of device with $W=1$ μm at 4.3 K, which is in diffusive regime, is well fitted to MS theory with the relationship of $d \propto (L_{GB})^{-1}$ in a two-dimensional system. Fig. 3b shows $L_{GB}$ dependence of $R_B$ at 4.3 K normalized with that at 285 K for devices of $W=150$ nm. As shown in Supplementary Fig. 6, for devices of $W=1$ μm, normalized $R_B$ shows similar $L_{GB}$ dependence of resistivity since the film is in a diffusive regime for all $L_{GB}$ values. For devices with $W=250$ nm (Supplementary Fig. 6b) and 150 nm, however, MS model fails to explain the normalized $R_B$ for small value of $L_{GB}$ where $l_{mfp}$ becomes longer than $W$ and $R_B$ becomes negative.

**Geometrical effect on the transport measurement for SCCF**

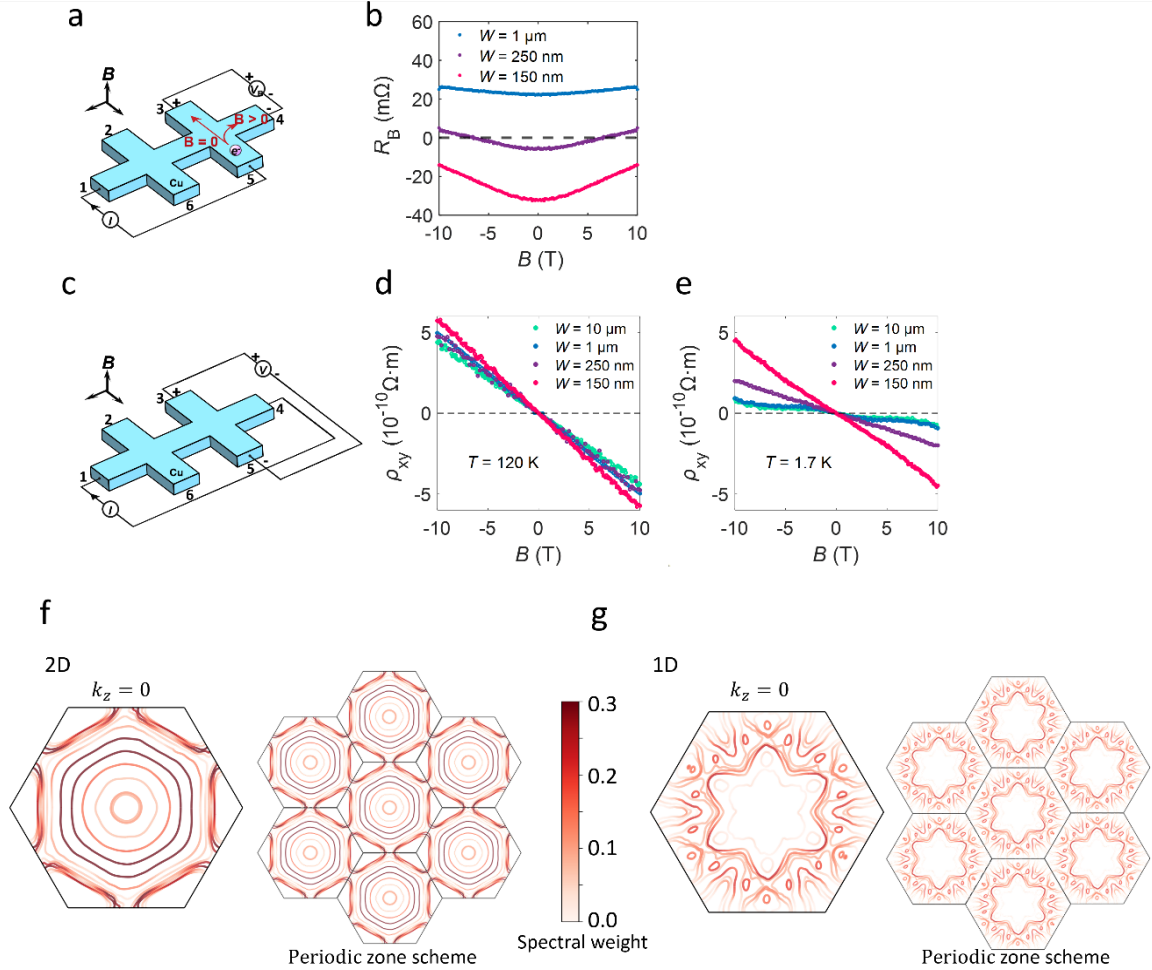

**Fig. 4 | Magnetic field dependence of bend resistance and geometrical effect. a,** Measurement schematics for symmetrized bend resistance ($R_B^s$) in magnetic field ($B$). **b,** $B$ dependence of $R_B^s$ in a single-crystalline Cu(111) thin film (SCCF) Hall bar with a width $W$=1 μm, 250 nm, and 150 nm at temperature $T$=4.3 K. **c,** Measurement schematics for Hall resistivity ($\rho_{xy}$). **d-e,** $B$ dependence of antisymmetrized Hall resistivity $\rho_{xy}^{as}$ in SCCF Hall bar with various $W$ at $T$=120 K (**d**) and $T$=1.7 K (**e**). **f,** Calculated effective Fermi surfaces of Cu(111) thin film in the 2D limit at $k_z = 0$. The inner surfaces of 2D limit Fermi surface are splitted bands due to confined geometry along the out-of-plane direction (left panel). Calculated effective Fermi surfaces of 2D Cu in periodic zone scheme (right panel). 2D effective Fermi surface is obtained by band unfolding to 3D. Color of the band represents the spectral weight of the effective band. **g,** Calculated effective Fermi surfaces of 1D Cu(111) at $k_z = 0$ (left panel) and calculated effetive Fermi surfaces of 1D Cu in periodic zone scheme (right panel). 1D effective Fermi surface is obtained by band unfolding to 2D.

To further study the ballistic transport behaviour under magnetic field, we investigated the magnetic field dependence of bend resistance $R_B$ (Fig. 4). We symmetrized $R_B$ to reduce intermixed signal from slight asymmetries and misalignments between contacts. Magnetic field $B$ was applied normal to the SCCF, which is parallel to the (111) direction. As the magnitude of $|B|$ increases, symmetrized bend resistance $R_B^s$ gradually rises due to the electron deflection by the Lorentz force resulting in fewer electrons reaching the opposite terminal 3 (Fig. 4a)[29]. This ballistic transport behaviour can be qualitatively explained using the Landauer-Büttiker approach[41]. $R_B^s$ can be expressed in terms of the forward transmission probability $P_{FW}$, transmission probabilities for turning left $P_L$ and turning right $P_R$, as shown in Eq. (4).

$$R_B^s = \frac{h}{e^2} \frac{P_L P_R - P_{FW}^2}{D} \quad (\text{Eq. 4a})$$

$$D = (P_L + P_R)[2P_{FW}(P_{FW} + P_L + P_R) + P_L^2 + P_R^2] \quad (\text{Eq. 4b})$$

For $B=0$, $P_{FW}$ is significantly larger than both $P_L$ and $P_R$. As a result, ($P_L P_R - P_{FW}^2$) becomes negative, leading to $R_B^s < 0$. When magnetic field is introduced, the ballistic electrons are deflected by Lorentz force with a reduction in $P_{FW}$ and an increase in either $P_L$ or $P_R$, resulting in an increase of $R_B^s$. As shown in Fig. 4b, the increase of $R_B^s$ as a function of $B$ is more pronounced in the ballistic regime ($W$=150 nm) compared to the diffusive regime ($W$=1 μm). This is because, in the ballistic regime, electron trajectories are more well-defined.

The antisymmetrized Hall resistivity $\rho_{xy}^{as}$ at 120 K linearly increases with $B$ as observed in typical metallic films (Fig. 4d), whereas at 1.7 K (Fig. 4e) $\rho_{xy}^{as}(B)$ becomes nonlinear and its slope decreases at wider devices. The difference with temperature could be due to changes in the Fermi surface topology as a result of temperature-induced chemical potential shift (Supplementary note 1). In a recent study, it was proposed that such nonlinear Hall effect in SCCF(111) is attributed to hole-carrier dominant transport supported by two-carrier model fitting analysis[33]. The nonlinear Hall effect at 1.7 K of the samples with $W$ = 10 μm and 1 μm widths agrees with the results observed in the recent study for thin films of thickness below 205 nm[33], which showed that in SCCFs with no grain boundaries, both electrons and holes exist simultaneously. The nonlinear Hall effect observed in devices with line widths of 10 μm and 1 μm (Figure 4e) suggests that electrons do not experience any confinement in their behavior due to the pattern size at the micrometer scale width of the 90-nm-thick thin film. However, when the width is reduced to 250 nm or below, the Hall effect becomes linear again

(Fig. 4e), which reflects that the carriers become electrons. To explain the phenomenon where the Hall effect reverts to a linear regime in 1D, we calculated and compared the effective band structures of a 2D Cu(111) thin film (Fig. 4f) along specific k-points with those of a 1D Cu(111) rod (Fig. 4g). In the band structure of the 2D thin film, both electron and hole orbits are distinctly visible. In contrast, the band structure of the 1D structure appears much more complex due to confined geometry in both the out-of-plane and in-plane directions.

The Fermi surface of the 2D thin film exhibits splitting of degenerate bands due to confined geometry along the [111] direction (Fig. 4f, left panel). In the representation of periodic zone scheme of Fermi surface (Fig. 4f, righ panel), the hole bands are more dominantly visible[33]. The Fermi surface of the 1D structure appears featureless and monotonous providing little physical insight. To extract meaningful insight, we calculated the effective Fermi surface obtained by band unfolding to 2D (Fig. 4g, left panel). In the unfolded representation of the periodic zone scheme (Fig. 4g, right panel), the hole bands are significantly reduced, while the electron orbits appear dominant. Although the system size for the calculation is much smaller than that of the actual device, the calculation shows the trend of the reduction of hole orbits as the system geometry approaches 1D regime. This qualitatively explains the experimental observation that the two-carrier model-induced nonlinear Hall effect (NHE) evolves into a linear Hall effect as $W$ decreases. For reference, Supplementary Fig. 7 compares the 3D, 2D, and 1D Fermi surfaces at given $k_z$.

**Conclusion**

In this study, we observed ballistic electronic transport in nano-devices based on Cu(111) thin film and investigated its dependence on the film quality. Hall-bar shaped devices were fabricated with atomically flat 90-nm-thick SCCF. We obtained negative bend resistance in the devices as direct evidence of ballistic transport. In a perpendicular magnetic field, ballistic electrons are disturbed, resulting in an increase of the bend resistance. Furthermore, we determined the distribution of GBs and TBs in our samples through EBSD measurement and show that the resistivity of Cu thin film devices is primarily dependent on GBs, but insensitive to TBs. Furthermore, we experimentally observed and theoretically explained that as the Cu(111) thin film transitions from the 2D to the 1D limit, the hole orbits observed in 2D disappear due to confined geometry in both the out-of-plane and in-plane directions, which suggests the transition from the nonlinear Hall effect driven by two carriers to a linear Hall effect.

Our findings not only provide a novel platform for investigating the intrinsic quantum mechanical properties of Cu but also have potential implications for advancing high-performance electronic[42] and spintronic devices by preserving quantum information, including momentum, quantum phase, and spin. They may help address critical reliability challenges in semiconductor technology, such as Joule heating[24] and electromigration[25] in Cu interconnects [23]. Moreover, recent discoveries reveal that the Fermi surface of Cu exhibits a topologically nontrivial genus[19], opening the door to various topology-related experiments in ballistic metals[19–21]].

# Methods

**Preparation of thin SCCF using the ASE technique.**

The ASE method involves stacking atoms individually without forming clusters, which can cause arbitrary deposition. To achieve this, we modified the conventional sputtering system as follows. The network of conducting wires, including cables, in the conventional sputtering system was upgraded with single-crystal (SC) copper wires. These wires were fabricated by cutting single crystal Cu wafers in a spiral fashion using wire electrical discharge machining (wire-EDM). The wafers themselves were sliced from an SC ingot grown by the Czochralski method. To minimize vibration caused by ambient noise, we implemented a mechanical noise reduction system. While such minute vibrations may not significantly affect conventional thin film growth, especially for polycrystalline films (PCCFs), they can cause irreversible stacking faults that disrupt the initial nucleation and lateral growth processes, particularly the coherent coplanar merging of nuclei. The thin film growth system creates a stable environment for single-atom deposition, aiming to achieve atomically flat surfaces through the precise stacking of single atoms.

The optimized sputtering conditions are as follows:

- Substrate: A double-sided polished (001) $Al_2O_3$ wafer with a thickness of 430 μm.
- Deposition Temperature: Approximately 170°C.
- RF Power: 13.56 MHz at 30 W.
- Target-to-Substrate Distance: 95 mm.
- Base Pressure: Maintained at less than $2 \times 10^{-7}$ Torr.
- Working Pressure: $5.4 \times 10^{-3}$ Torr with an Ar gas (99.9999% (6N)) flow of 50 sccm.

The relationship between the deposition time and the thickness of the thin film (or the average growth rate) was determined from the average deposition time of a 200-nm-thick film grown under optimal conditions.

**Sample nano-fabrication and transport measurement**

To fabricate the Hall bar devices using the prepared SCCF, we partially removed sections of the single-crystalline Cu via sequential e-beam lithography and Ar ion milling, resulting in the formation of Hall bars with varying linewidths (Supplementary Fig. 2). As shown in Fig. 1b, the Cu atomic spacing consistently measures 2.07 Å along the (111) direction, confirming the pristine crystalline quality of the SCCF, even after the patterning processes.  Temperature dependence of bend resistance and magnetic field dependence of Hall resistivity of SCCF Hall bar devices were measured in Oxford Instrument Teslatron PT with a base temperature of 1.7 K connected with low-pass RC filters. Grain boundary dependence of bend resistance, magnetic field dependence of bend resistance and longitudinal resistance of SCCF Hall bar devices were measured in Oxford Instrument Heliox with a base temperature of 4.3 K connected with low-pass RC filters. Bend resistance and longitudinal resistance were measured by the DC IV-sweep method by measuring the voltage drop with a Keithley 2000 voltmeter, and the current was applied to 200 nA through Yokogawa GS610 source measure unit with a load resistance of 100 kΩ. Hall resistance was measured with delta measurement method by measuring the voltage drop with low-noise Keithley 2182 nanovoltmeter, and current with 1 mA was applied through low-noise Keithley 6221 current source.

**Structural information of Hall bar pattern**

To investigate the cross-sectional structure of ballistic transport in SCCF, the ADF imaging mode of an aberration-corrected STEM instrument (JEM-ARM200CF, JEOL) operating at 300 kV was employed. The angle range of the ADF detector was set to 45-175mrad, and the semi-convergence angle of the condenser lens was ≈24 mrad. In combination with STEM imaging, elemental mapping of the Cu films was performed in the same STEM image mode using an EEL spectroscopy (Quantum ER965, Gatan). Cross-sectional TEM sampling was conducted using the Ga ion mailing and slicing method in an FIB scanning electron microscope (Helios NanoLab 450, Thermo Fisher Scientific). Low-energy Ar ion beam milling at 700 V for 10 min was sequentially performed as a post-surface treatment to remove the damaged surface layer that usually forms during heavy Ga ion beam milling.

**Theoretical Calculations**

All total energy calculations and electronic band structure calculations were performed based on first-principles density functional theory[Kohn65] as implemented by Kresse and Joubert[Kres99] using the projector augmented-wave method[Bloc94]. The exchange–correlation functional was modelled using the generalized gradient approximation in the Perdew–Burke–Ernzerhof form[Perd96]. All calculations were spin-polarized, and the positions of the atoms and the size and shape of the unit cell were fully relaxed to obtain the optimized lattice structure. All the atoms of the bulk Cu were fully relaxed until the force on the atom was less than 0.001 eV/Å and the change in total energy was less than $10^{-6}$ eV. The electron wavefunctions were expanded via a plane-wave basis set with a cut-off energy of

400 eV for both the bulk, slab, and rod calculations. The 12-ML Cu(111) slab structure was used to simulate the 2D limit thin film while 6-ML×6-ML rod struture was used for 1D limit nano-rod. We maintained a 20 Å vacuum to prevent interactions between the periodic images.

## Data availability

The authors declare that the main data supporting the findings of this study are available within the article and its Supplementary Information files.

## Acknowledgements

This research was supported by the National Research Foundation of Korea (NRF) (nos., RS-2022-NR068223, RS-2024-00393599, RS-2024-00442710, RS-2024-00444725 received by G.-H.L., 2021R1A5A1032937, RS-2024-00406152 and RS-2024-00455226 received by S-Y.J.) and by the ITRC (Information Technology Research Center) support program (IITP-2025-RS-2022-00164799 received by G.-H.L.). G.-H.L. acknowledges the support of the Samsung. Electronics Co., Ltd (IO201207-07801-01) and S.-Y.J. acknowledges the support of the Samsung Science and Technology Foundation (project number SRFC-MA2202-02). Y.-M.K. acknowledges use of the TEM instrument supported by the Advanced Facility Center for Quantum Technology at SKKU.

## Author contributions

G.-H.L., S.-Y.J., H.-J.L. and S.-G.K. conceived this study. Y.J.C. performed the transport measurements. S.-Y.J., S.J.K. and Y.L. performed the Cu thin film growth and observation using AFM, XRD, EBSD, and SEM. Y.H.K. and Y.-M.K. performed TEM measurements and analyses. S.G.K. and H.J.L. supervised the work. G.-H.L., S.-Y.J. and Y.J.C. wrote the manuscript. All authors participated in the manuscript review.

## Competing interests

The authors have no competing interests.

## Additional information

**Supplementary information** The online version contains supplementary material available at https://doi.org./

**Correspondence** and requests for materials should be addressed to Se-Young Jeong or Gil-Ho Lee

# Supplementary Information

## Ballistic transport in nanodevices based on single-crystalline Cu thin films

Yongjin Cho[1], Su Jae Kim[2], Min-Hyoung Jung[3], Yousil Lee[4], Hu Young Jeong[5], Young-Min Kim[3], Hu-Jong Lee[1], Seong-Gon Kim[6]\*, Se-Young Jeong[7,8,9]\*, and Gil-Ho Lee[1,10]\*

[1]Department of Physics, Pohang University of Science and Technology, Pohang, Republic of Korea

[2]Crystal Bank Research Institute, Pusan National University, Busan 46241, Republic of Korea

[3]Department of Energy Science, Sungkyunkwan University, Suwon 16419, Republic of Korea

[4]Copper Innovative Technology (CIT) Co. Busan 46285, Republic of Korea

[5] Graduate School of Semiconductor Materials and Devices Engineering, Ulsan National Institute of Science and Technology, Ulsan 44919, Republic of Korea

[6]Department of Physics and Astronomy, Mississippi State University, Mississippi State, MS 39762, USA

[7]Department of Physics, Korea Advanced Institute of Science and Technology (KAIST), Daejeon 34141, Korea

[8]Gordon Center for Medical Imaging, Department of Radiology, Massachusetts General Hospital and Harvard Medical School, Boston, MA 02114, USA

[9]Department of Optics and Mechatronics Engineering, Engineering Research Center for Color-Modulated Extra-Sensory Perception Technology, Pusan National University, Busan 46241, Republic of Korea

[10]Asia Pacific Center for Theoretical Physics, Pohang 37673, Republic of Korea

\*Corresponding authors. E-mail:
sk162@msstate.edu (S.-G.K.);
sjeong9@kaist.ac.kr (S.-Y.J);
lghman@postech.ac.kr (G.-H.L)

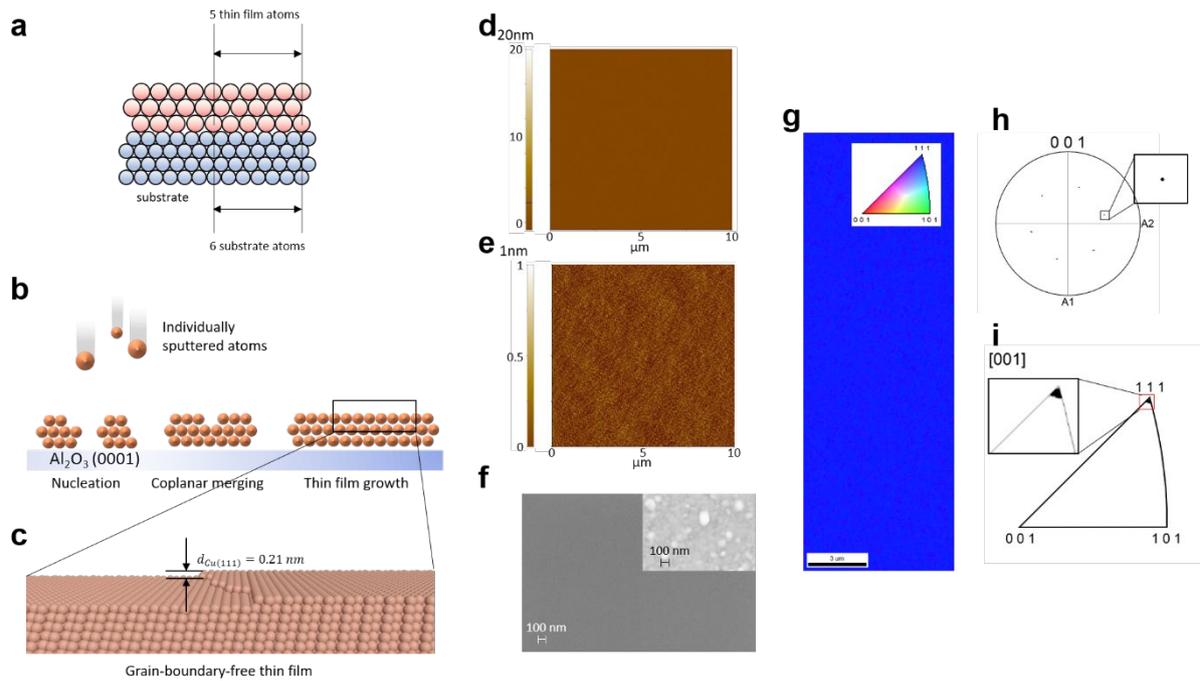

**Supplementary Fig. 1 | Single-crystal thin film used for ballistic transport measurements.**
**a,** Scheme of EADM at the heteroepitaxy. **b,** 3 initial stages of thin film growth: 1. nucleation stage, 2. coherent-coplanar merging stage and 3. layer-by-layer growth stage. **c,** Scheme of surface and atomic stacking of SCCF. **d-e,** Actual atomic force microscopy images of samples presented with 20 nm resolution (d) and 1 nm resolution (e). **f.** SEM image of SCCF surface. Inset: an SEM image of a polycrystalline sample for comparison. **g,** EBSD map showing ideal crystallinity with alignment along the (111) plane. **h,** Inverse pole figure displaying a single spot corresponding to the (111) plane. **i,** [100] pole figure revealing the six-fold symmetry of {100}.

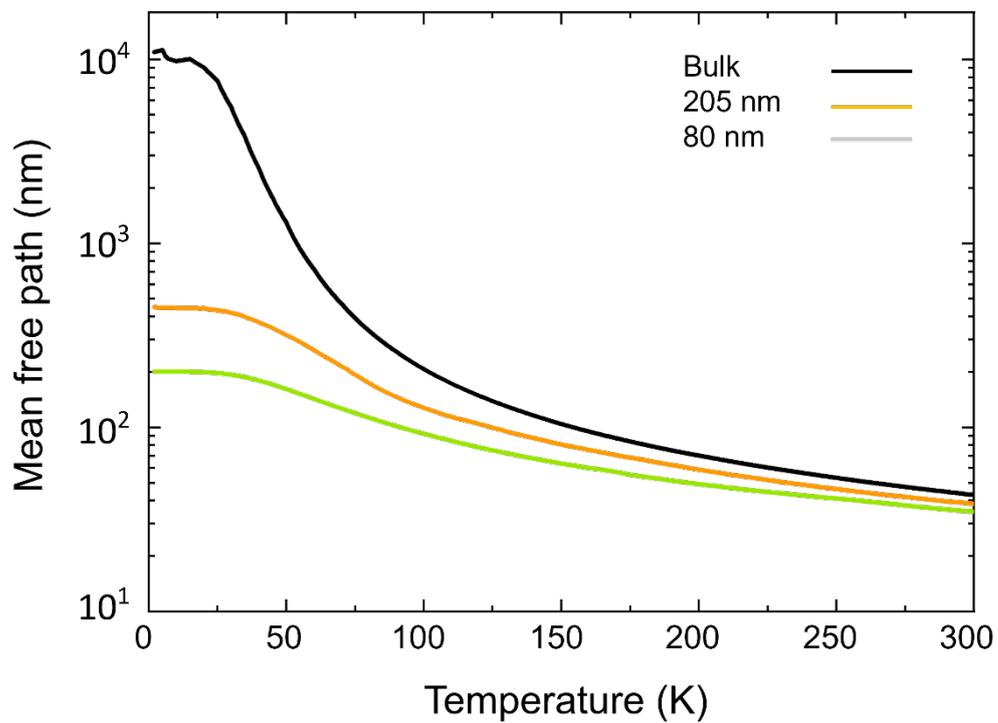

**Supplementary Fig. 2**. Mean free paths of electrons in single-crystal bulk and thin films with thicknesses of 205 and 80 nm, respectively. The mean free path was calculated from the relationship between the resistivity and mean free path.

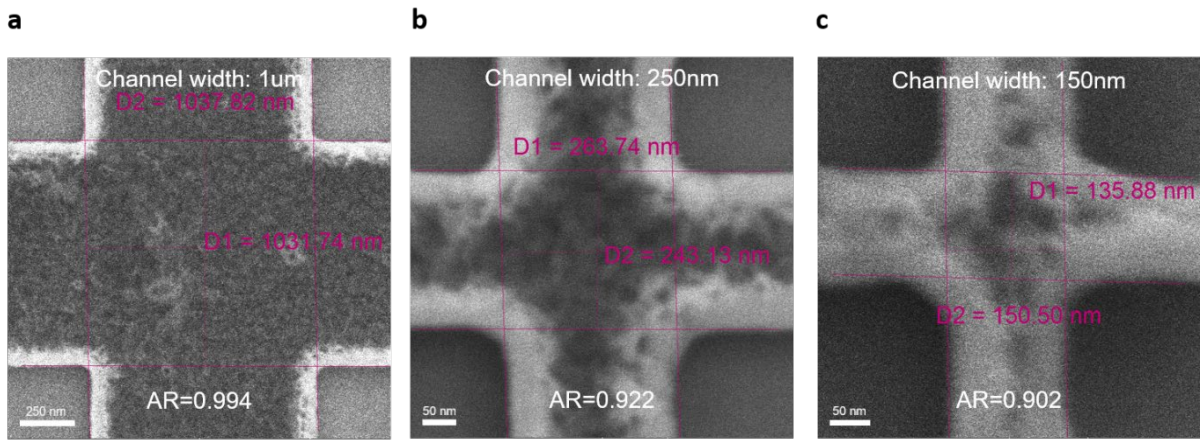

**Supplementary Fig. 3 | SEM image of SCCF Hall bars with various channel widths. a,** SEM image of an SCCF Hall bar device with a channel width $W$ = 1 μm. (**b**), $W$ = 250 nm (**c**) and $W$ = 150 nm

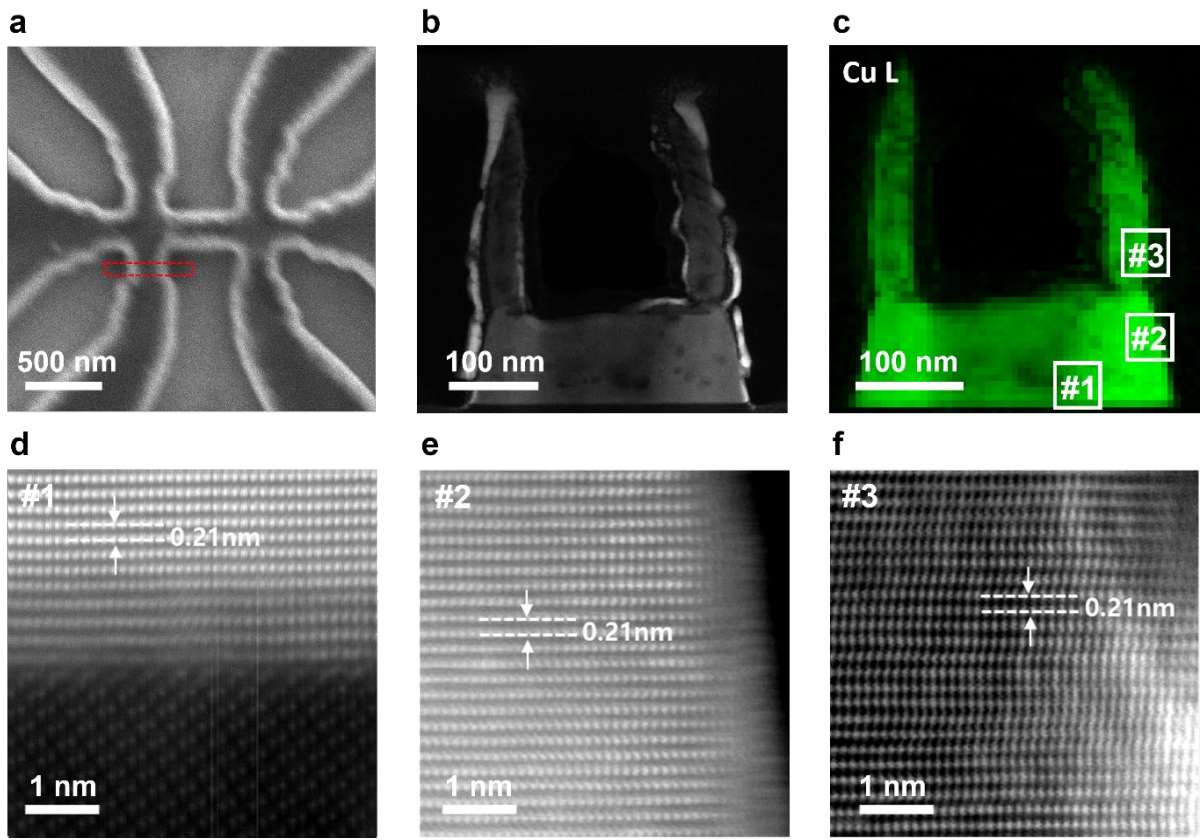

**Supplementary Fig. 4 | Cross-sectional TEM image of ballistic transport in SCCF. a,** SEM image of an SCCF Hall bar device with a channel width (*W*) of 250 nm. **b,** Low-magnification ADF-STEM image displaying the cross-sectional view of the SCCF Hall bar device. **c,** EELS-based elemental map of Cu. **d–f,** Enlarged atomic structure images for the regions marked by white boxes in **c**.

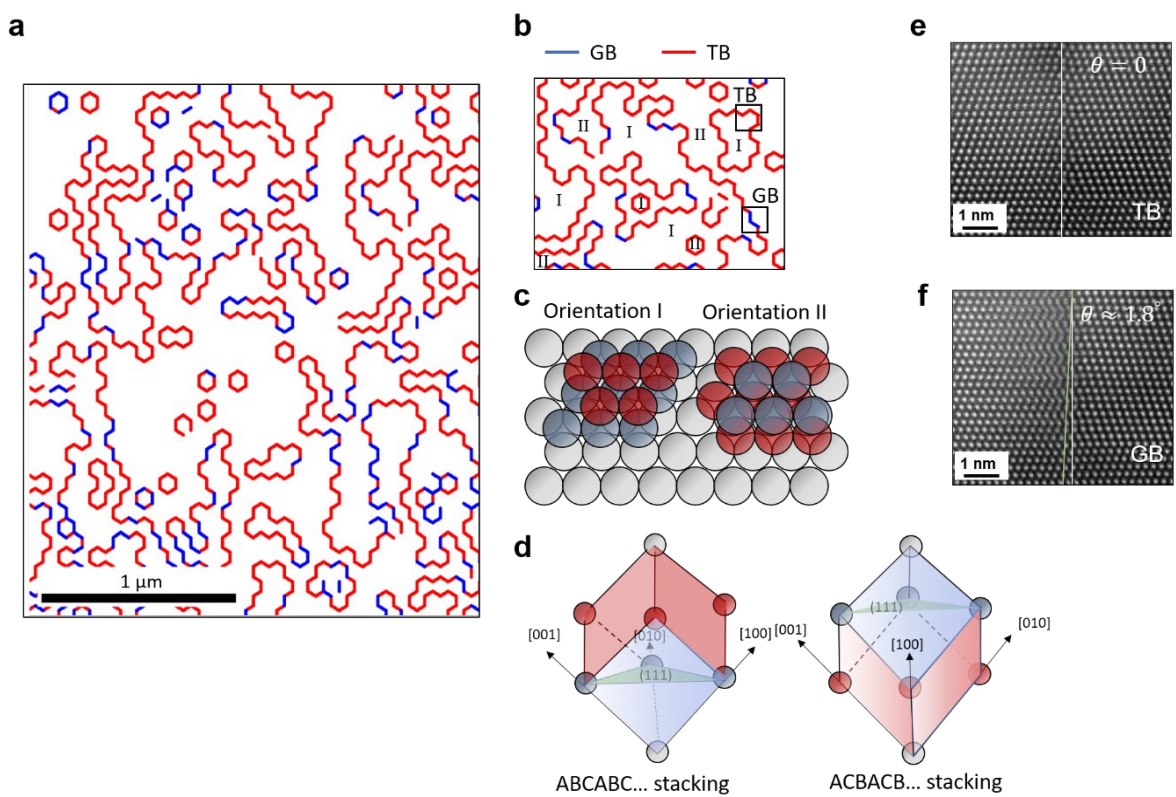

**Supplementary Fig. 5 | Crystallographic structure at GB and TB. a,** Misorientation line map of a single crystalline Cu thin film almost aligned along Cu(111), but still with a few GBs. **b**, Enlarged images of **a** with marks of orientations I and II. **c**, Atomic stacking of orientations I and II following ABC… and ACB..., respectively. **d**, 3D crystal structures of two different orientations I and II stacked along the diagonal direction of the cubic system. **e–f,** High-angle annular dark-field scanning transmission electron microscopy (HAADF-STEM) images showing the atomic stacking order near the TB (e) and GB (f) marked by boxes in **b**.

**Supplementary note to Supplementary Fig. 5.**

We analyzed the data from single-crystalline Cu thin films to distinguish between SCCFs and single-crystalline Cu thin films that are close to SCCFs but still have GBs. The GBs in single-crystalline Cu thin films differ significantly from those in PCCFs. While PCCFs can have arbitrary orientations, orientations in single-crystalline thin films are limited to two possibilities. Even if a single-crystalline thin film contains some GBs, these are very close to the TBs with a slight misorientation of less than 2° vertically. Nevertheless, electrons scatter at these slightly misoriented GBs, making ballistic transport difficult to observe. Therefore, ballistic behavior is only observed in extremely clean SCCF systems without GBs.

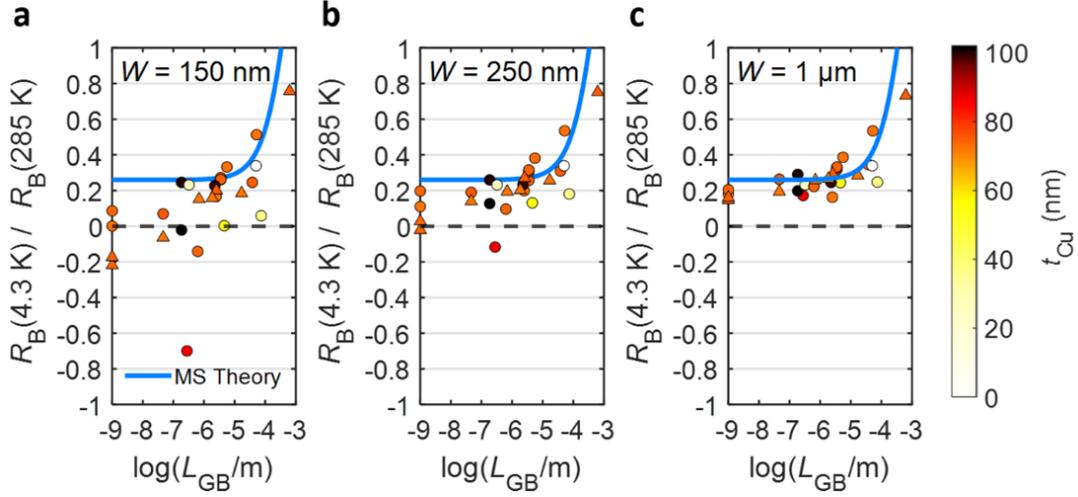

**Supplementary Fig. 6 | GB dependence of electrical transport. a-c,** Normalized bend resistance $R_B$ as a function of GB length $L_{GB}$ for $W = 1$ μm (**a**), $W = 250$ nm (**b**) and $W = 150$ nm (**c**). Blue solid line represents the MS theory fit for device with $W = 1$ μm, circles and triangles represent the device without and with annealing processes, respectively; the symbol colors represent thickness, $t_{Cu}$.

**Supplementary note to Supplementary Fig. 6.**

Supplementary Figs. 6a–c show the $L_{GB}$ dependence of $R_B$ at 4.3 K, normalized with that at 285 K. For devices with $W = 1$ μm, the normalized $R_B$ shows similar $L_{GB}$ dependence of resistivity as the film is in the diffusive regime for all $L_{GB}$ values. However, for devices with $W = 250$ and 150 nm, the MS model fails to explain the normalized $R_B$ for small values of $L_{GB}$, where $l_{mfp}$ exceeds $W$, and $R_B$ becomes negative.

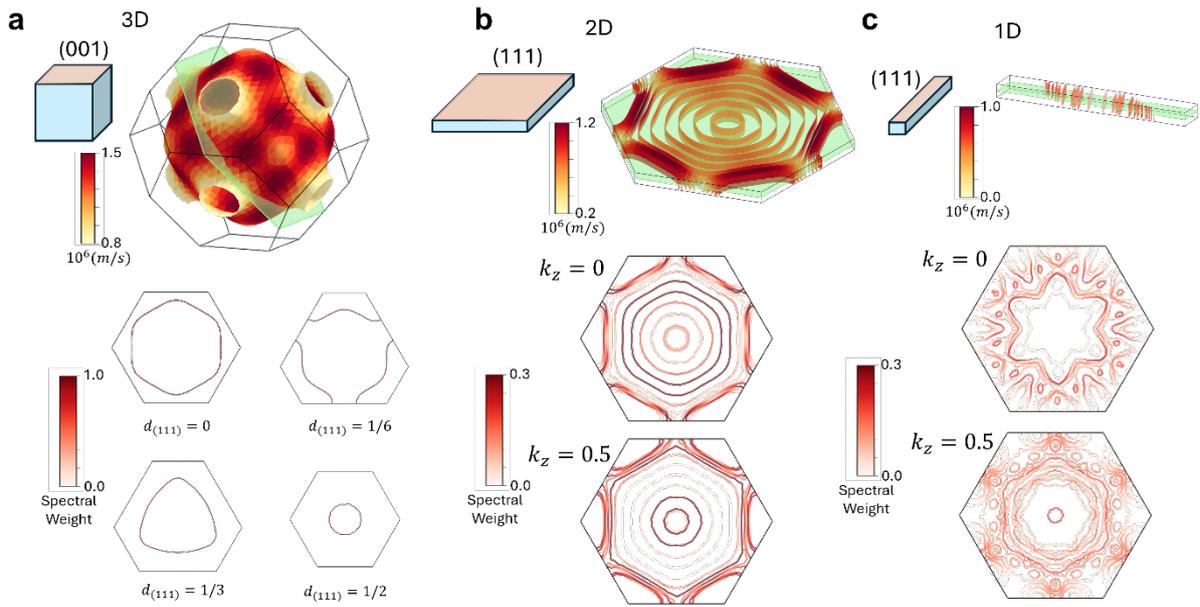

**Supplementary Fig. 7 | Comparison of Fermi surfaces of 3D, 2D and 1D Cu (111). a,** Fermi surface of Cu bulk (upper panel) and the intersections of Fermi surfaces with the (111) slice plane at various distances from the zone centre. **b,** Fermi surface of 2D Cu thin film (upper panel) and the 2D effective Fermi surfaces unfolded to 3D at different $k_z$ values. **c,** Fermi surface of 1D Cu(111) rod (upper panel) and the 1D effective Fermi surfaces obtained by band unfolding at different $k_z$ values. Color on the 3D Fermi surface represents the magnitude of the group velocity of the electrons. Color of the band in Fermi slices represents the spectral weight of the effective band.

**Supplementary note to Figs. 4d and e.**

If the density of state $g(\varepsilon)$ changes significantly near the Fermi energy $\varepsilon_F$, then temperature-induced chemical-potential $\mu(T)$ shift occurs owing to thermal excitation of the electrons, as shown in equation (5)[S1,S2].

$$\mu(T) = \varepsilon_F - \frac{\pi^2}{6}(k_B T)^2 \frac{g'(\varepsilon_F)}{g(\varepsilon_F)} \quad (5)$$

The shift in chemical potential due to temperature changes is negligible in typical metals such as bulk copper. However, in semimetals, where conduction band minimum and valence band maximum are near the chemical potential, significant shifts can occur with temperature changes owing to considerable variations in the density of the state close to the Fermi level. This result suggests that the electronic band structure of the 90-nm-thick SCCF, modified by quantum confinement effect perpendicular to the film plane, has semimetal-like properties. This results in a change of carrier type at low temperature from single electron carrier to two carriers (electron and hole) due to temperature-induced chemical potential shift[25,S1-S5]. For widths <250 nm, the Hall resistivity at 1.7 K (Fig. 4e) becomes linear, consistent with the phenomenon observed at 120 K. In addition, the semimetal-like behavior disappears possibly because of additional quantum confinement effects along the channel width direction.